\documentclass[doublecol]{epl2}
\usepackage[T1]{fontenc}
\usepackage[latin9]{inputenc}
\usepackage{amsmath}
\usepackage{esint}

\makeatletter
\@ifundefined{textcolor}{}
{%
 \definecolor{BLACK}{gray}{0}
 \definecolor{WHITE}{gray}{1}
 \definecolor{RED}{rgb}{1,0,0}
 \definecolor{GREEN}{rgb}{0,1,0}
 \definecolor{BLUE}{rgb}{0,0,1}
 \definecolor{CYAN}{cmyk}{1,0,0,0}
 \definecolor{MAGENTA}{cmyk}{0,1,0,0}
 \definecolor{YELLOW}{cmyk}{0,0,1,0}
 }

\title{The suppression of Curie temperature by Sr doping in diluted ferromagnetic semiconductor (La$_{1-x}$Sr$_{x}$)(Zn$_{1-y}$Mn$_{y}$)AsO}
\shorttitle{The suppression of $T_C$ by Sr doping in DFS
(La$_{1-x}$Sr$_{x}$)(Zn$_{1-y}$Mn$_{y}$)AsO}
\author{Cui Ding\inst{1}, Xin Gong\inst{1}, Huiyuan Man\inst{1}, Guoxiang Zhi\inst{1},  Shengli Guo\inst{1}, Yang Zhao\inst{2,3}, Hangdong Wang\inst{4}, Bin Chen\inst{4} and F.L. Ning\inst{1}\footnote{Electronic address: ningfl@zju.edu.cn}}
\shortauthor{Cui Ding\etal}

 \institute{
  \inst{1} Department of Physics, Zhejiang University - Hangzhou 310027, China\\
  \inst{2} NIST center for Neutron Research, National Institute of Standards and Technology, Gaithersburg, Maryland 20899, USA\\
  \inst{3} Department of Materials Science and Engineering, University of Maryland, College Park, Maryland 20742,
  USA\\
  \inst{4} Department of Physics, Hangzhou Normal University - Hangzhou 310016, China
}


\abstract{ (La$_{1-x}$Sr$_{x}$)(Zn$_{1-y}$Mn$_{y}$)AsO is a two
dimensional diluted ferromagnetic semiconductor that has the
advantage of decoupled charge and spin doping. The substitution of
Sr$^{2+}$ for La$^{3+}$ and Mn$^{2+}$ for Zn$^{2+}$ into the parent
semiconductor LaZnAsO introduces hole carriers and spins,
respectively. This advantage enables us to investigate the influence
of carrier doping on the ferromagnetic ordered state through the
control of Sr concentrations in
(La$_{1-x}$Sr$_x$)(Zn$_{0.9}$Mn$_{0.1}$)AsO. 10 \% Sr doping results
in a ferromagnetic ordering below $T_C$ $\sim$ 30 K. Increasing Sr
concentration up to 30 \% heavily suppresses the Curie temperature
and saturation moments. Neutron scattering measurements indicate
that no structural transition occurs for
(La$_{0.9}$Sr$_{0.1}$)(Zn$_{0.9}$Mn$_{0.1}$)AsO below 300 K.}

\begin{document}

\maketitle

\section{Introduction}

The observation of ferromagnetic ordering in III-V (Ga,Mn)As below
Curie temperature $T_C$ $\sim$ 60 K by Ohno et al \cite{ohno} has
generated extensive research into the diluted ferromagnetic
semiconductors (DFS). After almost two decades of efforts, $T_C$ has
been improved to as high as 200 K with Mn doping level of $\sim$ 12
\% \cite{Wang, Zhaojianhua,Zhaojianhua2,DietlOhon}. This temperature
is still far below room temperature which is the prerequisite for
practical application of spintronics \cite{Zutic}. Improving $T_C$
in homogenous (Ga,Mn)As thin films is one of the objectives in the
research of DFS. On the other hand, understanding the mechanism of
ferromagnetic ordering is hindered by some inherent difficulties. In
(Ga,Mn)As, the substitution of Mn$^{2+}$ for Ga$^{3+}$ provides not
only local moments but also hole carriers. It is generally believed
that ferromagnetic ordering can arise only when spins are
effectively mediated by carriers \cite{Dietl}. However, during the
fabrication of (Ga,Mn)As thin films, some Mn atoms enter
interstitial sites and behave as a double doner, which make it
difficult to determine precisely the amount of Mn that substitute Ga
at ionic sites. Seeking for new ferromagnetic semiconductor systems
with more controllable charge and spin densities might be helpful to
understand the general mechanism of ferromagnetism in DFS.

Recently, several bulk DFS systems that are derivatives of Fe-based
high temperature superconductor have been reported. The first
Fe-based superconductor is 1111-type oxypnictides,
LaFeAs(O$_{1-x}$F$_{x}$) \cite{Kamihara}, which has a
superconducting transition temperature $T_c$ = 26 K. With identical
two dimensional crystal structure, three 1111 type DFS systems,
(La,Ba)(Zn,Mn)AsO with $T_C$ $\sim$ 40 K \cite{Ding1},
(La,Ca)(Zn,Mn)SbO with $T_C$ $\sim$ 40 K \cite{IOP1111},
(La,Sr)(Cu,Mn)SO \cite{Yang} with $T_C$ $\sim$ 210 K have been
reported. Similarly, two bulk form DFS systems,
(Ba,K)(Zn,Mn)$_2$As$_2$ \cite{Zhao} with $T_C$ $\sim$ 180 K and
(Ba,K)(Cd,Mn)$_{2}$As$_{2}$ with $T_C$ $\sim$ 17 K\cite{Yang2}, have
been reported. These two systems are structurally identical to that
of 122 type iron pnictides superconductor, (Ba,K)Fe$_2$As$_2$ ($T_c$
= 38 K)\cite{Rotter}. The third DFS family reported recently is
Li(Zn,Mn)Pn (Pn = P, As)\cite{Deng1,Deng2} with $T_C$ $\sim$ 50 K,
which are fabricated by doping Mn into the I-II-V direct gap
semiconductors LiZnPn (Pn = P, As). LiZnAs can also be viewed as a
derivative of the third family of Fe-based superconductors LiFeAs
($T_c$ = 18 K)\cite{WangXC}. The fourth family of Fe-based
superconductors is 11 type FeSe$_{1+\delta}$ ($T_c$ = 8
K)\cite{Hsu}, which can be paralleled to the well investigated II-VI
DFS, i.e., (Zn,Mn)Se. There are two more families of Fe-based
superconductors, namely, 32522 type
(Ca$_3$Al$_2$O$_{5}$)Fe$_2$As$_2$ ($T_c$$\sim$ 30.2 K)\cite{Shirage}
and 42622 type Sr$_4$V$_2$O$_6$Fe$_2$As$_2$(T$_c$$\sim$37.2
K)\cite{Zhu}. Very recently, 32522 type DFS
Sr$_3$La$_2$O$_5$(Zn,Mn)$_2$As$_2$ with Curie temperature $T_C$
$\sim$ 40 K \cite{Man} and 42622 type DFS
Sr$_4$Ti$_2$O$_6$(Zn,Mn)$_2$As$_2$ with $T_C$ $\sim$ 25K have been
reported\cite{Wang42622}.

Different to thin film form (Ga,Mn)As specimens, above new DFS
systems are all in bulk form. The availability of a specimen in bulk
form enables the investigation of DFS by powerful magnetic probes
including NMR (Nuclear Magnetic Resonance), $\mu$SR (muon Spin
Rotation) and neutron scattering. NMR investigation of I-II-V DFS
Li(Zn,Mn)P by Ding et al. has shown that the spin-lattice relaxation
rate $\frac{1}{T_{1}}$ of Li(0) site (zero means that no Mn atoms at
N.N. (nearest neighbor) Zn site of Li) exhibits a kink around $T_C$,
which indicates that Li(0) sites are indeed under the influence of
ferromagnetic Mn spin fluctuations \cite{DingNMR}. Furthermore,
$\frac{1}{T_{1}}$ of Li(Mn) site is temperature independent above
$T_C$, i.e., $\frac{1}{T_{1}}$ $\sim$ 400 s$^{-1}$, indicating that
Mn spin-spin interaction extends over many unit cells with an
interaction energy scale $\mid$J$\mid$ $\sim$ 100 K. On the other
hand, $\mu$SR measurements have shown that bulk form I-II-V, 1111
and 122 DFSs all share a common ferromagnetic mechanism as that of
(Ga,Mn)As thin film \cite{Deng1, Ding1, Zhao}.

Another important feature of the newly fabricated bulk DFSs is that
they all have advantages of decoupled carriers and spins doping.
Here Mn$^{2+}$ substitution for Zn$^{2+}$ introduces only spins, and
carriers are introduced at a different site. Only when both carriers
and spins are introduced simultaneously, can ferromagnetic ordering
develop\cite{Ding1}. In this paper, we dope Sr and Mn into the
direct gap parent semiconductor LaZnAsO up to the doping level of 30
\%. We found that chemical solubility is 20 \%, which is much higher
than 10\% of doping Ba. This allows us to investigate the high
doping regime in a more reliable manner. The Curie temperature $T_C$
in (La$_{1-x}$Sr$_{x}$)(Zn$_{1-x}$Mn$_{x}$)AsO increases from 30 K
of $x$ = 0.10 to 35 K of $x$ = 0.20, but decreases to 27 K for $x$ =
0.30. For a fixed Mn concentration of $x$ = 0.10, we found that more
hole doping than 10 \% suppresses both $T_C$ and the saturation
moments. 30 \% hole doping suppresses saturation moment by almost an
order, and no strong enhancement in magnetization curve that
corresponds to the ferromagnetic ordering has been observed.

\section{Experiments}

Polycrystalline specimens of
(La$_{1-x}$Sr$_{x}$)(Zn$_{1-y}$Mn$_{y}$)AsO were prepared through
solid state reaction method by mixing intermediate products LaAs,
ZnAs, MnAs, ZnO, MnO and SrO with nominal concentrations. The
mixture was then made into pellets and heated to 1150 $^{\circ}$C
slowly. It was kept at 1150 $^{\circ}$C for 40 hours before cooling
down with shutting off furnace. The intermidiate products LaAs, ZnAs
and MnAs were prior produced with mixing high purity elements La,
Zn, Mn and As and heating at 900 $^{\circ}$C for 10 hour. The
processes of mixing were carried in high purity Ar atmosphere in a
glove box and the mixtures were sealed in an evacuated silica tube
before heating. The polycrystals were characterized by X-ray powder
diffraction and dc magnetization by SQUID (Superconducting Quantum
Interference Device). The electrical resistance was measured on
sintered pellets with typical four-probe method. Neutron scattering
measurements were performed at NIST Center for Neutron Research
(NCNR) using BT-1 powder Diffractometer.

\section{Results and discussion}

\begin{figure}[!htpb]  \centering
\centering
\includegraphics[width=3in]{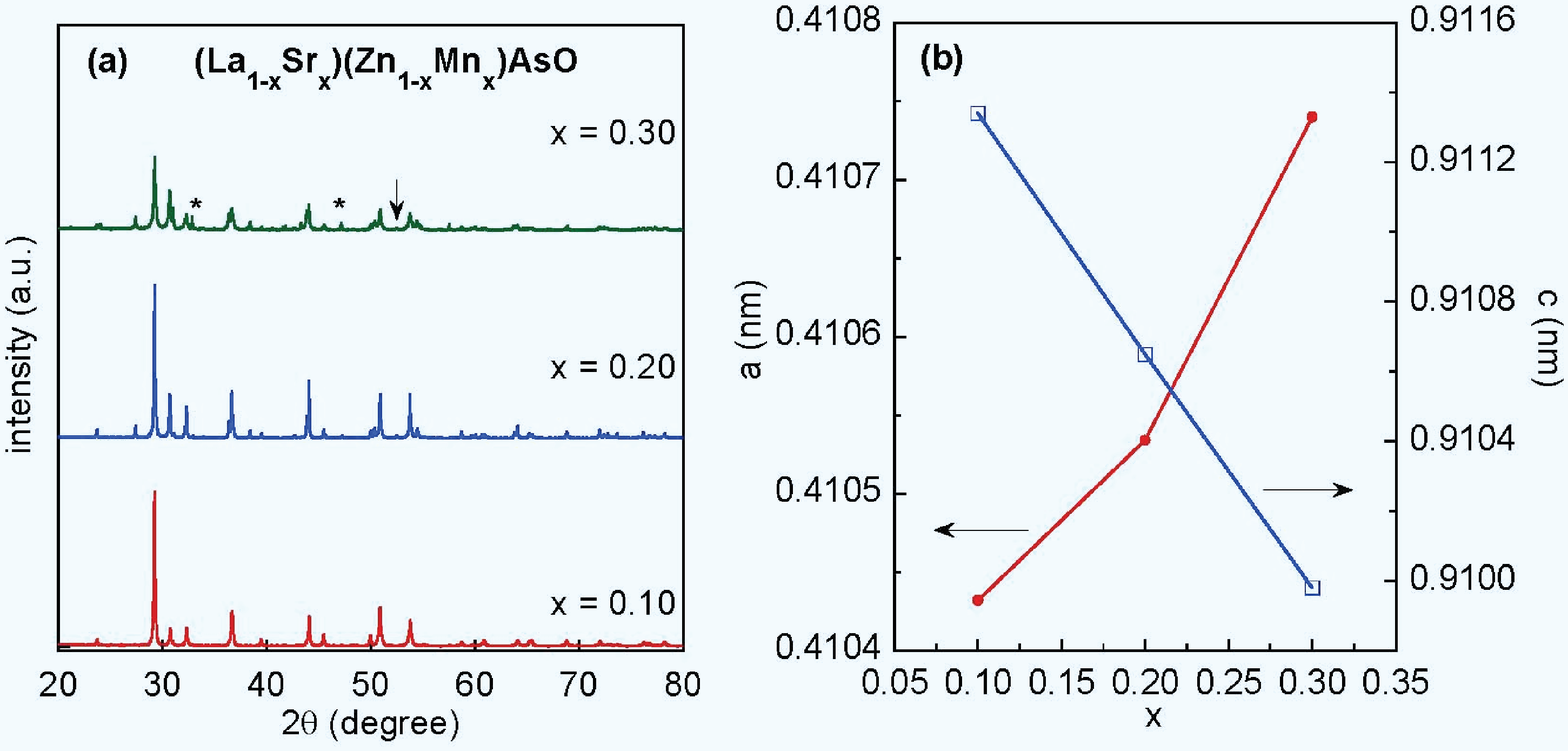}\vspace*{0cm}
\caption{(Color online) (a) Powder X-ray diffraction pattern of
(La$_{1-x}$Sr$_{x}$)(Zn$_{1-y}$Mn$_{y}$4)AsO ($x$ = $y$ = 0.10,
0.20, 0.30). Traces of La$_2$O$_3$ ($\downarrow$) and ZnAs$_2$
($\ast$) impurities are marked for $x$ = 0.30.
 (b)Lattice parameters for a axis (red solid circles) and c axis (blue open squares).} \label{Fig.1}
\end{figure}

In Fig. 1(a), we show the powder X-ray diffraction patterns of
(La$_{1-x}$Sr$_{x}$)(Zn$_{1-y}$Mn$_{y}$)AsO with Sr and Mn of equal
doping levels ($x$ = $y$ = 0.10, 0.20, 0.30). As demonstrated by the
Reitveld refinement for LaZnAsO compound \cite{Ding1}, the Bragg
peaks of the specimens can be well indexed into a tetragonal crystal
structure of space group P4/nmm. The lattice parameters for three
concentrations are shown in figure 1(b). The lattice constant $a$
monotonically increases while $c$ monotonically decreases with Sr
and Mn doping, indicating successful solid solution of (La,Sr) and
(Zn,Mn). We find the chemical solubility is as high as 20 \%, much
higher than the case of Ba and Mn doping into LaZnAsO which is only
10 \% \cite{Ding1}. This can be attributed to the much closer atomic
radius of La$^{3+}$ (0.106 nm) and Sr$^{2+}$ (0.113 nm). Traces of
La$_2$O$_3$ ($\downarrow$) and ZnAs$_2$ ($\ast$) impurities are
observed for $x$ = 0.30 concentration. These impurities are
non-magnetic, which do not affect our discussion below.

We show results of electrical resistance measured for parent
compound LaZnAsO, 10 \% Sr doped specimen
(La$_{0.9}$Sr$_{0.1}$)ZnAsO, and 10 \% Sr and Mn doped specimen
(La$_{0.9}$Sr$_{0.1}$)(Zn$_{0.9}$Mn$_{0.1}$)AsO in Fig. 2. The
parent compound is a direact-gap semiconductor with a band gap of
$\sim$ 1.5 eV \cite{Kayanuma}, and it displays a typical
semiconducting behavior. 10 \% Sr substitution for La readily
changes the semiconductor into a metal, as demonstrated by the
decreasing resistivity with temperature decreasing. More
interestingly, once 10 \% Mn are doped into
(La$_{0.9}$Sr$_{0.1}$)ZnAsO, the specimen returns to a
semiconducting behavior. It seems that spins induced by Mn atoms
localize the hole carriers. We do a tentative fitting of the
$\rho$(T) data for (La$_{0.9}$Sr$_{0.1}$)(Zn$_{0.9}$Mn$_{0.1}$)AsO
near room temperature with a thermally activated conductivity,
$\frac{1}{\rho}=Cexp(-E_a/2k_BT)$. The fitting is good, and the
scenario of the disorder induced localization mechanism seems not
applicable here. We also find an activation energy of 0.056 eV,
which is much smaller than the gap energy of LaZnAsO.

\begin{figure}[!htpb] \centering
\centering
\includegraphics[width=3in]{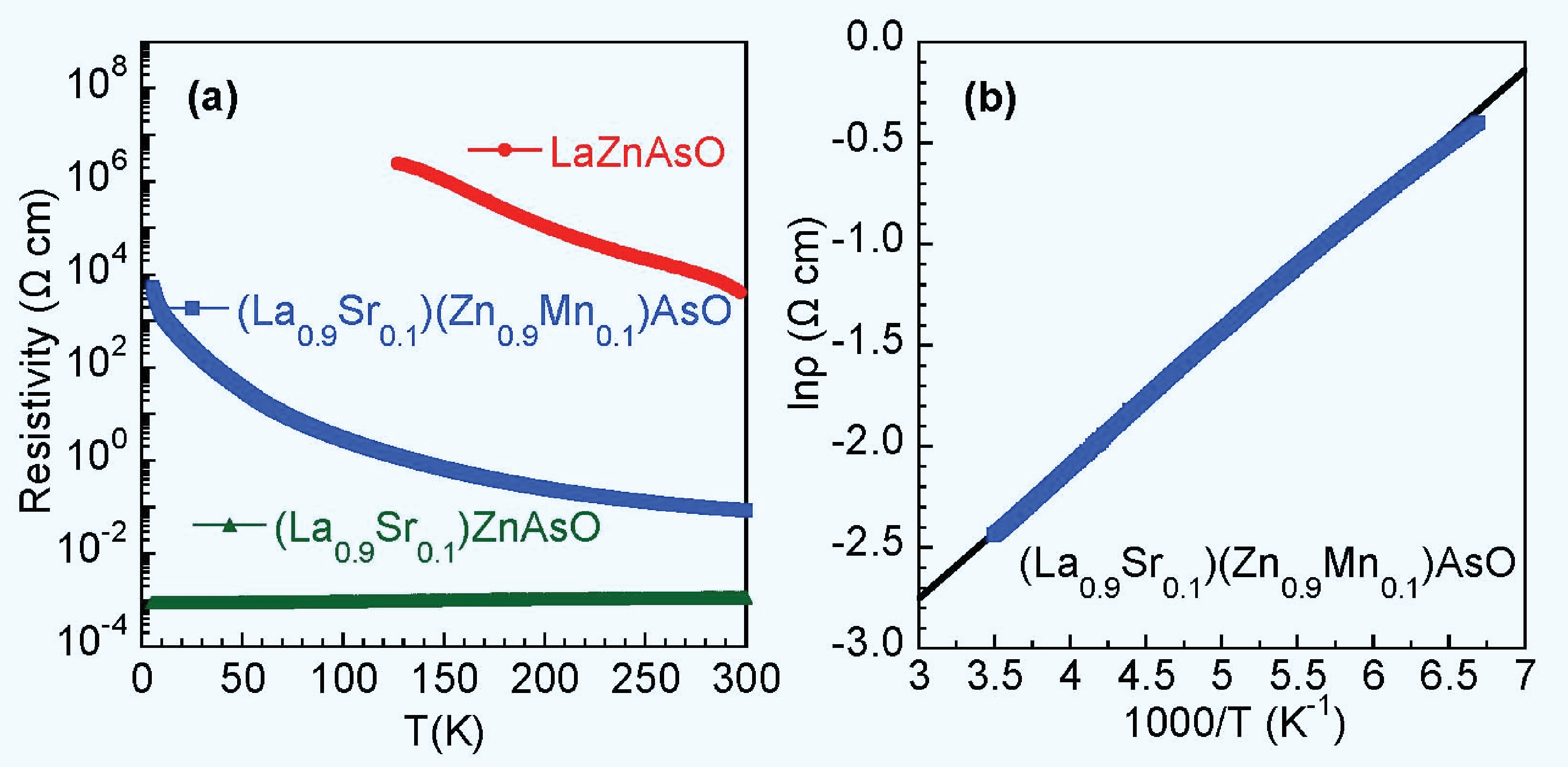}\vspace*{0cm}
\caption{(Color online) (a) The electrical resistivity for parent
compound LaZnAsO, 10 \% Sr doped specimen
(La$_{0.9}$Sr$_{0.1}$)ZnAsO, and 10 \% Sr and Mn doped specimen
(La$_{0.9}$Sr$_{0.1}$)(Zn$_{0.9}$Mn$_{0.1}$)AsO. (b) The fitting of
resistivity for (La$_{0.9}$Sr$_{0.1}$)(Zn$_{0.9}$Mn$_{0.1}$)AsO with
an activation funtion.} \label{Fig.2}
\end{figure}

\begin{figure}[!htpb] \centering
\centering \vspace*{0cm}
\includegraphics[width=3in]{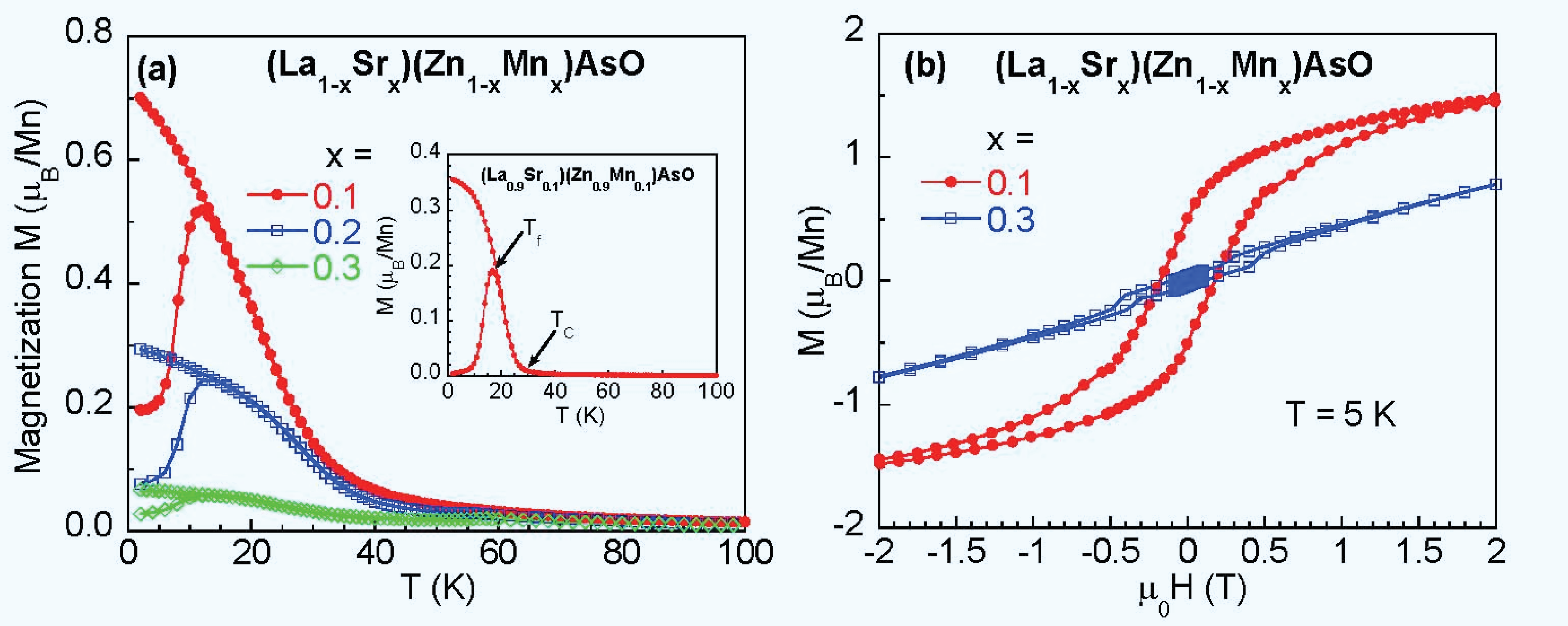}\vspace*{0cm}
\caption{(Color online)(a)dc magnetization M measured under ZFC and
FC condition for (La$_{1-x}$Sr$_x$)(Zn$_{1-y}$Mn$_y$)AsO($x$ = $y$ =
0.10, 0.20, 0.30) at $\mu_0$H = 0.1 Tesla; the inset shows M of
(La$_{0.9}$Sr$_{0.1}$)(Zn$_{0.9}$Mn$_{0.1}$)AsO measured at $\mu_0$H
= 0.005 Tesla, $T_C$ and $T_f$ are marked by arrows. (b)The
isothermal magnetization measurements for
(La$_{1-x}$Sr$_x$)(Zn$_{1-y}$Mn$_y$)AsO($x$ = $y$ = 0.10, 0.30) at 5
K. } \label{Fig.3}
\end{figure}

In Fig. 3(a), we show dc magnetization curve of
(La$_{1-x}$Sr$_{x}$)(Zn$_{1-y}$Mn$_{y}$)AsO ($x$ = $y$ = 0.10, 0.20,
0.30) measured with 0.1 Tesla external field under zero-field-cooled
(ZFC) and field-cooled (FC) condition (note that Sr atoms and Mn are
in the same doping level). We define the Curie temperature of
ferromagnetic ordering, $T_C$, as the temperature where
magnetization curve measured at 0.005 Tesla displays a sharp
enhancement, as marked by the arrow in the inset of Fig. 3(a). As
the doping level increases, $T_C$ increases from 30 K for $x$ = 0.10
to 35 K for $x$ = 0.20, and then decreases to 27 K for $x$ = 0.30.
The saturation moment is 0.71 $\mu$$_B$/Mn for $x$ = 0.10, which is
comparable to that of
(La$_{0.9}$Ba$_{0.1}$)(Zn$_{0.9}$Mn$_{0.1}$)AsO\cite{Ding1}. The
saturation moment is suppressed monotonically to 0.07 $\mu$$_B$/Mn
at the doping level of $x$ = 0.30. Another feature of magnetization
curve is that ZFC and FC curves split at a temperature below $T_C$,
as marked by the arrow in inset of Fig.3(a). This temperature is
defined as $T_f$, which is the onset temperature of static spin
freezing, as has been investigated by $\mu$SR measurement of
(La,Ba)(Zn,Mn)AsO\cite{Ding1}. $T_f$ decreases monotonically from 17
K ($x$ = 0.10) to 12 K ($x$ = 0.30).

\begin{figure}[!htpb]  \centering
\centering \vspace*{0cm}
\includegraphics[width=3in]{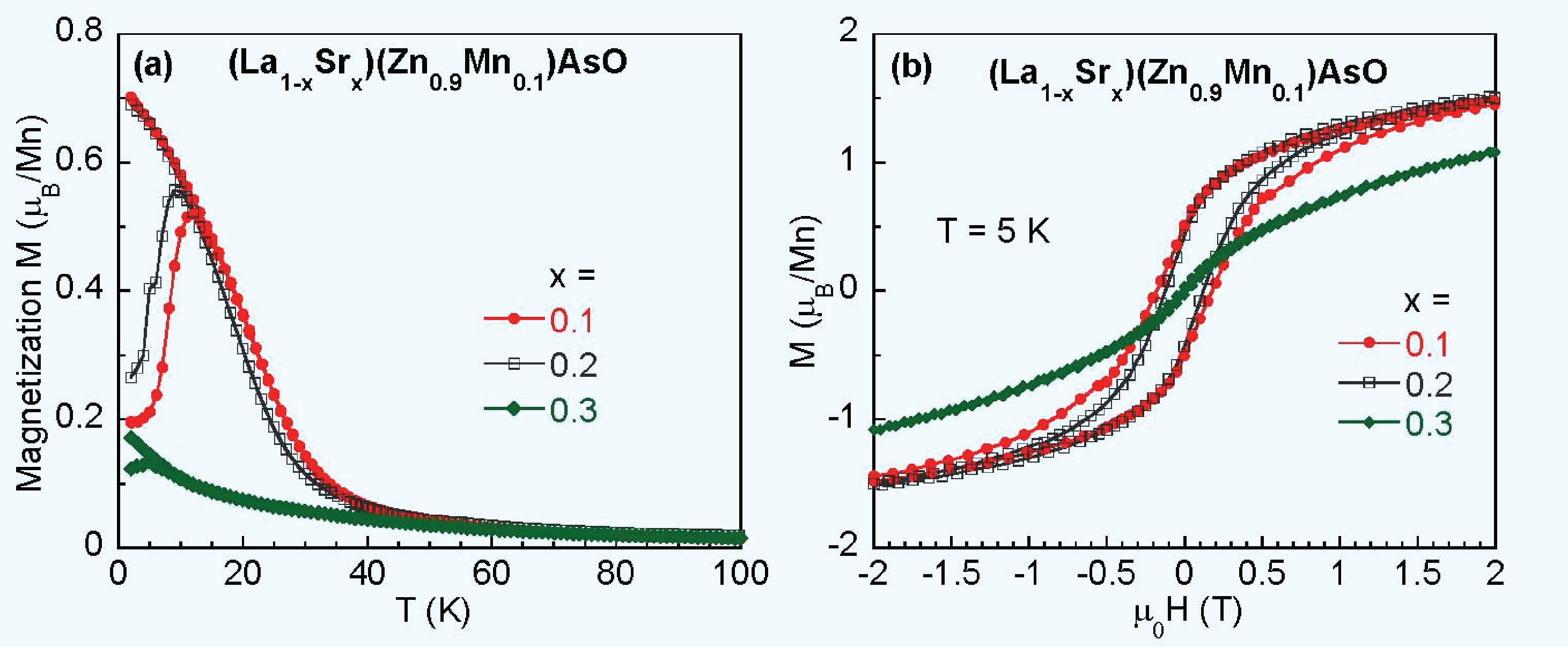}\vspace*{0cm}
\caption{(Color online)(a)dc magnetization M measured under ZFC and
FC condition for (L$a_{1-x}$Sr$_x$)(Zn$_{0.9}$Mn$_{0.1}$)AsO ($x$ =
0.10, 0.20, 0.30) at $\mu_0$H = 0.1 Tesla. (b) The isothermal
magnetization measurements for
(L$a_{1-x}$Sr$_x$)(Zn$_{0.9}$Mn$_{0.1}$)AsO ($x$ = 0.10, 0.20, 0.30)
at 5 K. } \label{Fig.4}
\end{figure}

The isothermal magnetization at 5 K for
(La$_{0.9}$Sr$_{0.1}$)(Zn$_{0.9}$Mn$_{0.1}$)AsO and
(La$_{0.7}$Sr$_{0.3}$)(Zn$_{0.7}$Mn$_{0.3}$)AsO is shown in figure
3(b). With the increasing of Sr and Mn concentrations, the coercive
fields are quickly suppressed from 0.178 Tesla to 0.102 Tesla. For
comparison, the coercive field of (Ga$_{0.965}$Mn$_{0.035}$)As is
$\sim$ 0.005 Tesla \cite{ohno}. On the other hand, the saturation
remanence of (La$_{0.9}$Sr$_{0.1}$)(Zn$_{0.9}$Mn$_{0.1}$)AsO is
$\sim$ 0.5 $\mu$$_B$/Mn, an order of magnitude larger than $\sim$
0.04 $\mu$$_B$/Mn of (La$_{0.7}$Sr$_{0.3}$)(Zn$_{0.7}$Mn$_{0.3}$)AsO
. The suppression of $T_C$, saturation moments, and coercive fields
with higher Mn doping levels is very likely arising from competition
of direct antiferromagnetic exchange interaction from N.N. Mn atoms
at Zn sites. The probability to find two Mn at N.N. Zn sites is
$P(N; x) = C_N^4x^N(1-x)^{(4-N)}$ = 29.16 \% for $x$ = 0.10 (where N
= 1 and $x$ = 0.10). This probability increases to 41.16 \% for the
doping level of $x$ = 0.30. The direct antiferromagnetic coupling
between Mn-Mn pairs eventually results in the antiferromagnetic
ordering in LaMnAsO at $T_N$ = 317 K\cite{AFM}.

\begin{figure}[!htpb]  \centering
\includegraphics[width=2.8in]{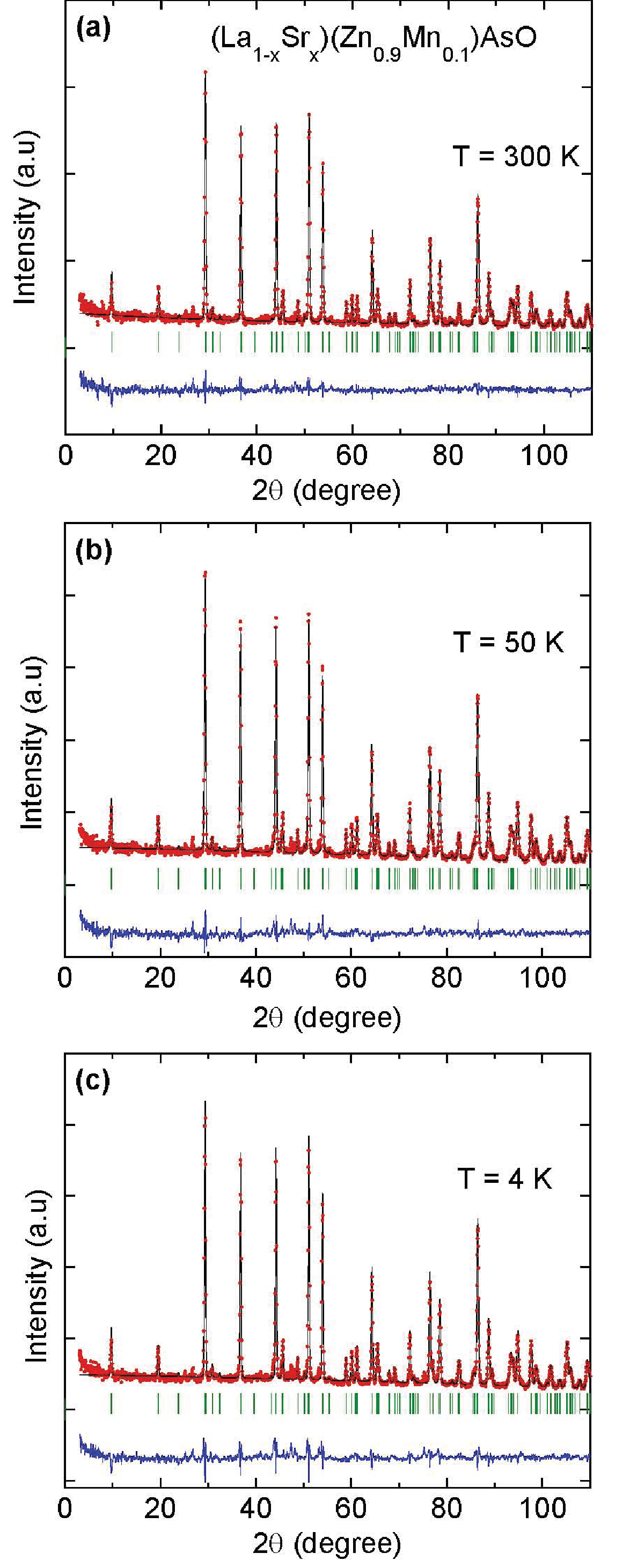}\vspace*{0cm}
\caption{Neutron diffraction pattern for the powder specimen of
(La$_{0.9}$Sr$_{0.1}$)(Zn$_{0.9}$Mn$_{0.1}$)AsO at 300K (a), 50K (b)
and 4K(c) at NIST. Error bars are smaller than plot symbols.}
\label{Fig.5}
\end{figure}

Since we can control spins and carriers separately, it will be
interesting to investigate the influence of different carrier doping
levels on the magnetic state at a fixed Mn concentration. We fix Mn
at the level of 0.10, and enhance the doping levels of Sr up to 20
\% and 30 \%. The ZFC and FC dc-magnetization curves of three
specimens are plotted in Fig. 4(a) and their isothermal
magnetization at 5 K are plotted in Fig. 4(b). Compared to 10 \% Sr
doping sample, both $T_C$ and saturation moments are only slightly
suppressed with 20 \% Sr doping. When Sr doping level is increased
to $x$ = 0.30, both $T_C$ and saturation moments are heavily
suppressed. The saturation moment decreases from 0.71 $\mu$$_B$/Mn
for $x$ = 0.10 to 0.17 $\mu$$_B$/Mn for $x$ = 0.30. Similarly, $T_f$
decreases from 17 K to 5 K and coercive field declines from 0.178
Tesla to 0.022 Tesla. The saturation remanence decreases from $\sim$
0.5 $\mu$$_B$/Mn for $x$ = 0.10 to 0.03 $\mu$$_B$/Mn for $x$ = 0.30.
In early stage of DFS research, it has been theoretically proposed
that spins are mediated by hole carriers through RKKY
interaction\cite{Dietl}. We can write the RKKY exchange interaction
as $J \sim \frac{cos(2k_Fr)}{r^3}$, where $k_F$ is the radius of
Fermi surface if assuming the Fermi surface is a spherical shape,
and $r$ the distance between two localized moments. The first
oscillation period of RKKY interaction supports ferromagnetic
coupling. In present work, doping more Sr into
(L$a_{1-x}$Sr$_x$)(Zn$_{0.9}$Mn$_{0.1}$)AsO has introduced extra
hole carriers, which modifies the density of states and the Fermi
surface, and subsequently the ferromagnetic ordering.

To investigate the spin structure, we conducted neutron diffraction
of the powder specimen
(La$_{0.9}$Sr$_{0.1}$)(Zn$_{0.9}$Mn$_{0.1}$)AsO (which has the
largest saturation moment size) at 4 K , 50 K and 300 K at NCNR, and
show the results in Fig. 5. We found that neutron powder diffraction
pattern is in line with X-rays diffraction pattern, and no
impurities are observed. This indicates that Mn atoms are indeed
substituted into Zn sites. We spent 8 hours for a 3 grams sample at
4 K, but still could not separate the magnetic diffraction peaks
from the structural ones. In
(La$_{0.9}$Sr$_{0.1}$)(Zn$_{0.9}$Mn$_{0.1}$)AsO, the average moment
size is only $\sim$ 0.07 $\mu$$_B$/Mn(Zn), which is smaller than the
limit of neutron resolution $\sim$ 0.1 $\mu$$_B$ with current
experiment configuration. None the less, we do not observe
structural phase transition from 300 K to 4 K, which cross $T_C$ =
30 K.

\section{Summary and Conclusion}

To summarize, we successfully synthesized a new DFS system
(La$_{1-x}$Sr$_{x}$)(Zn$_{1-x}$Mn$_{x}$)AsO with doping level up to
30 \%. Maximum $T_C$ is as high as 35 K at the doping of $x$ = 0.20.
With the advantage of decoupled spin and carrier doping, we could
investigate the influence of carrier concentration on the
ferromagnetic ordering. We found that 30 \% Sr substitution for La
in (La$_{1-x}$Sr$_{x}$)(Zn$_{0.9}$Mn$_{0.1}$)AsO suppresses the
ferromagnetic ordering, leaving a spin glass like magnetic ordered
state. As we have shown in (LaBa)(ZnMn)AsO \cite{Ding1}, spins can
not order ferromagnetically without carrier doping. Our experimental
evidence shown in this work unequivocally demonstrates that too much
carriers suppresses both the Curie temperature and the saturation
moments. In other words, too much carriers are detrimental to the
ferromagnetic ordering as well. It requires a delicate balance
between carriers and spins to achieve highest $T_C$. In addition,
our neutron scattering experiments rule out the possibility of a
structural transition between 300 K and 4 K for
(La$_{0.9}$Sr$_{0.1}$)(Zn$_{0.9}$Mn$_{0.1}$)AsO specimen. Finally,
as stated previously, the common two dimensional crystal structure
shared by ferromagnetic (La$_{1-x}$Sr$_{x}$)(Zn$_{1-y}$Mn$_{y}$)AsO,
1111-type Fe-based high temperature superconductors and
antiferromagnetic LaMnAsO, makes it possible to make various
junctions through As layers.

\acknowledgments

The work at Zhejiang was supported by National Basic Research
Program of China (No.2014CB921203, 2011CBA00103), NSF of China (No.
11274268). F.L. Ning acknowledges helpful discussions with I. Mazin,
I. Zutic, Y.J. Uemura and C.Q. Jin.

\end{document}